\documentclass[prl,twocolumn,unsortedaddress,superscriptaddress]{revtex4-1}
\usepackage[english]{babel}
\usepackage[utf8]{inputenc}
\usepackage{graphicx}
\usepackage{comment}
\usepackage{hyperref}
\usepackage{multirow}
\usepackage{color}
\usepackage{amsmath,amssymb,amsfonts} 

\usepackage{xy}
\xyoption{matrix}
\xyoption{frame}
\xyoption{arrow}
\xyoption{arc}

\usepackage{ifpdf}
\ifpdf
\else
\PackageWarningNoLine{Qcircuit}{Qcircuit is loading in Postscript mode.  The Xy-pic options ps and dvips will be loaded.  If you wish to use other Postscript drivers for Xy-pic, you must modify the code in Qcircuit.tex}
\xyoption{ps}
\xyoption{dvips}
\fi

\entrymodifiers={!C\entrybox}

\newcommand{\ket}[1]{{\left\vert{#1}\right\rangle}}
\newcommand{\qw}[1][-1]{\ar @{-} [0,#1]}
\newcommand{\qwx}[1][-1]{\ar @{-} [#1,0]}
\newcommand{\cw}[1][-1]{\ar @{=} [0,#1]}

\newcommand{\gate}[1]{*+<.6em>{#1} \POS ="i","i"+UR;"i"+UL **\dir{-};"i"+DL **\dir{-};"i"+DR **\dir{-};"i"+UR **\dir{-},"i" \qw}
\newcommand{\meter}{*=<1.8em,1.4em>{\xy ="j","j"-<.778em,.322em>;{"j"+<.778em,-.322em> \ellipse ur,_{}},"j"-<0em,.4em>;p+<.5em,.9em> **\dir{-},"j"+<2.2em,2.2em>*{},"j"-<2.2em,2.2em>*{} \endxy} \POS ="i","i"+UR;"i"+UL **\dir{-};"i"+DL **\dir{-};"i"+DR **\dir{-};"i"+UR **\dir{-},"i" \qw}

\newcommand{\control}{*!<0em,.025em>-=-<.2em>{\bullet}}

\newcommand{\controlm}{*-<.0em,.0em>{\xy *=<.40em>!<0em,-.02em>[O][F]{}\POS!C\endxy}}
\newcommand{\ctrlm}[1]{\controlm \qwx[#1] \qw}

\newcommand{\ctrl}[1]{\control \qwx[#1] \qw}

\newcommand{\targ}{*+<.02em,.02em>{\xy ="i","i"-<.39em,0em>;"i"+<.39em,0em> **\dir{-}, "i"-<0em,.39em>;"i"+<0em,.39em> **\dir{-},"i"*\xycircle<.4em>{} \endxy} \qw}
\newcommand{\qswap}{*=<0em>{\times} \qw}
\newcommand{\multigate}[2]{*+<1em,.9em>{\hphantom{#2}} \POS [0,0]="i",[0,0].[#1,0]="e",!C *{#2},"e"+UR;"e"+UL **\dir{-};"e"+DL **\dir{-};"e"+DR **\dir{-};"e"+UR **\dir{-},"i" \qw}
\newcommand{\ghost}[1]{*+<1em,.9em>{\hphantom{#1}} \qw}

\newcommand{\gategroup}[6]{\POS"#1,#2"."#3,#2"."#1,#4"."#3,#4"!C*+<#5>\frm{#6}}

\newcommand{\lstick}[1]{*!R!<.5em,0em>=<0em>{#1}}

\newcommand{\Qcircuit}{\xymatrix @*=<0em>}

\renewcommand{\tabcolsep}{0.1cm}
\renewcommand{\arraystretch}{0.7}
\begin{document}

\title{Relativistic quantum chemistry on quantum computers}

\author{Libor Veis}
\email{libor.veis@jh-inst.cas.cz}
\affiliation{J. Heyrovsk\'{y} Institute of Physical Chemistry, ASCR, 18223 Prague, Czech Republic}
\affiliation{\mbox{Department of Physical and Macromolecular Chemistry, Charles University, 12840 Prague, Czech Republic}}

\author{Jakub Vi\v{s}\v{n}\'{a}k}
\affiliation{J. Heyrovsk\'{y} Institute of Physical Chemistry, ASCR, 18223 Prague, Czech Republic}

\author{Timo Fleig}
\affiliation{Laboratoire de Chimie et Physique Quantiques, Universit\'{e} Toulouse 3, IRSAMC, F-31062 Toulouse, France}

\author{Stefan Knecht}
\affiliation{Department of Physics and Chemistry, University of Southern Denmark, DK-5230 Odense M, Denmark}

\author{Trond Saue}
\affiliation{Laboratoire de Chimie et Physique Quantiques, Universit\'{e} Toulouse 3, IRSAMC, F-31062 Toulouse, France}

\author{Lucas Visscher}
\affiliation{Amsterdam Center for Multiscale Modeling, VU University Amsterdam, NL-1081 HV Amsterdam, Netherlands}

\author{Ji\v{r}\'{i} Pittner}
\email{jiri.pittner@jh-inst.cas.cz, corresponding author}
\affiliation{J. Heyrovsk\'{y} Institute of Physical Chemistry, ASCR, 18223 Prague, Czech Republic}

\date{\today}

\begin{abstract}
Last years witnessed a remarkable interest in application of quantum computing for solving problems in quantum chemistry more efficiently than classical computers allow. Very recently, even first proof-of-principle experimental realizations have been reported. However, so far only the non-relativistic regime (i.e. Schroedinger equation) has been explored, while it is well known that relativistic effects can be very important in chemistry. In this communication we present  the first quantum algorithm for relativistic computations of molecular energies. We show how to efficiently solve the eigenproblem of the Dirac-Coulomb Hamiltonian on a quantum computer and demonstrate the functionality of the proposed procedure by numerical simulations of computations of the spin-orbit splitting in the SbH molecule. Finally, we propose quantum circuits with 3 qubits and 9 or 10 CNOTs, which implement a proof-of-principle relativistic quantum chemical calculation for this molecule and might be suitable for an experimental realization.
\end{abstract}

\maketitle

Quantum computing \cite{nielsen_chuang} is one of the fastest growing fields of computer science nowadays. Recent huge interest in this interdisciplinary field has been fostered by the prospects of solving certain types of problems more effectively than in the classical setting \cite{shor_1994, grover_1997}. The prominent example is the integer factorization problem where quantum computing offers an exponential speedup over its classical counterpart \cite{shor_1994}. But it is not only cryptography that can benefit from quantum computers. As was first proposed by R. Feynman \cite{feynman_1982}, quantum computers could in principle be used for \textit{efficient} simulation of another quantum system. This idea, which employs mapping of the Hilbert space of a studied system onto the Hilbert space of a register of quantum bits (qubits), both of them being exponentially large, can in fact be adopted also in quantum chemistry.

Several papers using this idea and dealing with the interconnection of quantum chemistry and quantum computing have appeared in recent years. These cover: calculations of thermal rate constants of chemical reactions \cite{lidar_1999}, \textit{non-relativistic} energy calculations \cite{aspuru-guzik_2005, wang_2008, whitfield_2010, veis_2010}, quantum chemical dynamics \cite{kassal_2008}, calculations of molecular properties \cite{kassal_2009}, initial state preparation \cite{wang_2009, ward_2009}, and also first proof-of-principle experimental realizations \cite{lanyon_2010, du_2010, li_2011, lu_2011}. An interested reader can find a comprehensive review in \cite{kassal_review}.

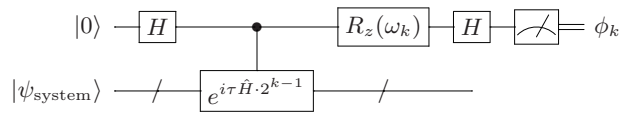
\begin{figure}[!ht]
 \begin{center}  
 \hskip 1cm 
 \mbox{
   \Qcircuit @C=1em @R=1em {
     \lstick{\ket{0}} & \gate{H} & \ctrl{1} & \gate{R_{z}(\omega_{k})} & \gate{H} & \meter & \cw & \phi_{k} \\
     \lstick{\ket{\psi_{\rm{system}}}} & {/} \qw & \gate{e^{i\tau\hat{H}\cdot2^{k-1}}} & {/} \qw & \qw
   }
 }
 \end{center}
 \caption{The $k$-th iteration of the iterative phase estimation algorithm (IPEA). The feedback angle $\omega_k$
depends on the previously measured bits.}
 \label{ipea_iteration} 
\end{figure}

An \textit{efficient} (polynomially scaling) algorithm for calculations of \textit{non-relativistic} molecular energies, that employs the phase estimation algorithm (PEA) of Abrams and Lloyd \cite{abrams_1999}, was proposed in the pioneering work by Aspuru-Guzik, et al. \cite{aspuru-guzik_2005}. When the ideas of measurement based quantum computing are adopted \cite{griffiths_1996}, the phase estimation algorithm can be formulated in an iterative manner [iterative phase estimation (IPEA)] with only one read-out qubit \cite{whitfield_2010,veis_2010}. If the phase $\phi$ ($0 \le \phi < 1$), which is directly related to the desired energy \cite{veis_2010}, is expressed in the binary form: $\phi = 0.\phi_1\phi_2\ldots$, $\phi_i = \{ 0 , 1 \}$, one bit of $\phi$ is measured on the read-out qubit at each iteration step. The algorithm is iterated backwards from the least significant bits of $\phi$ to the most significant ones, where the $k$-th iteration is shown in Figure \ref{ipea_iteration}. Not to confuse the reader, $\hat{H}$ in the exponential denotes the Hamiltonian operator, whereas $H$ (in a box) denotes the standard single-qubit Hadamard gate. $\ket{\psi_{\rm{system}}}$ represents the part of a quantum register that encodes the wave function of a studied system, $R_z$ is a $z$-rotation gate whose angle $\omega_k$ depends on the results of the previously measured bits \cite{veis_2010,whitfield_2010}, and parameter $\tau$ ensures that $0 \le \phi < 1$. The PEA always needs an initial guess of the wave function corresponding to the desired energy. This can be either the result of some approximate, polynomially scaling \textit{ab initio} method \cite{wang_2008, veis_2010}, or as originally proposed by Aspuru-Guzik, et al. \cite{aspuru-guzik_2005} the exact state or its approximation prepared by the adiabatic state preparation (ASP) method.

It is a well known fact that an accurate description of molecules with heavy elements requires adequate treatment 
of relativistic effects \cite{hess_2000}. The most rigorous approach [besides the quantum electrodynamics (QED) which is presently not feasible for quantum chemical purposes] is the four component (4c) no-pair formalism. Our work is based on the 4c electronic Dirac-Coulomb Hamiltonian (DCH) in the form

\vskip -0.3cm
\begin{equation}
  \hat{H} = \sum_{i=1}^N \left [c(\boldsymbol\alpha_i \cdot \mathbf{p}_i) + \beta^{\prime}_i m c^2 - \phi_{nuc} \right]
+ \sum_{i<j} \frac{1}{r_{ij}} +V_{NN}.
  \label{dirac_coulomb}
\end{equation}

\noindent
Dirac matrices appearing in the one-electron part are defined as
 
\begin{equation}
\boldsymbol{\alpha} = \left( \begin{array}{cc}  0 &\boldsymbol{\sigma} \\ \boldsymbol{\sigma} &0\end{array} \right)\quad\mbox{and}\quad\beta =  \left( \begin{array}{c c}  I_2 &0 \\ 0& -I_2\end{array} \right), \quad \beta^{\prime} = \beta - I_{4},
\end{equation}

\noindent
the former in terms of the Pauli spin matrices $\boldsymbol{\sigma}$. The DCH is known to cover the major part of the spin-orbit interaction and also scalar relativistic effects. Using this type of Hamiltonian represents no loss of generality for our purposes, since a transition to the Dirac-Coulomb-Breit Hamiltonian \cite{dyall} and the inclusion of the corresponding integrals
requires a classically polynomial effort.

We also adopt the no-pair approximation (NPA),
widely used in relativistic quantum chemistry \cite{dyall}, in which the N-particle basis of Slater determinants is constructed from positive-energy bispinors only. 
For a more detailed discussion about the DCH and approximations employed in the relativistic quantum chemistry, see the Supplementary Information.

The use of a 4c relativistic formalism brings in three major computational difficulties compared to the non-relativistic case: 
(1) working with 4c orbitals (bispinors), (2) complex algebra when molecular symmetry is low, and (3) rather large Hamiltonian matrix eigenvalue problems [due to larger mixing of states than in the non-relativistic (NR) case]. The central objective of this work is to address these problems in regard of an application of a quantum computer and the extension of the quantum full configuration interaction (qFCI) method to the relativistic regime. 

We will start the description of the algorithm with a mapping of the relativistic quantum chemical wave function onto a quantum register. The simplest (scalable) NR approach, the direct mapping (DM) \cite{aspuru-guzik_2005}, assigns each spin orbital one qubit ($\ket{0}$ = unoccupied,  $\ket{1}$ = occupied). The relativistic case is similar due to the NPA.  Moreover, because of the time-reversal symmetry of the Dirac equation, bispinors occur in degenerate Kramers pairs \cite{dyall} denoted $A$ and $B$ (analogy to $\alpha$ and $\beta$ spin in NR treatment) and the relativistic DM thus looks like: one qubit for bispinor $A$ and one for $B$. The 4c character of molecular bispinors therefore does not complicate the approach substantially [note that as in the NR case, the Hartree-Fock (HF) calculation is done on a classical computer and only the exponentially scaling FCI on a quantum one].

The DM is known to be not optimal as it maps the whole Fock space of the system on the Hilbert space of qubits. For this reason, compact mappings from a subspace of fixed-electron-number and spin- or symmetry-adapted wave functions have been proposed \cite{aspuru-guzik_2005, wang_2008}. However, general factorization schemes [i.e. algorithms to systematically generate quantum circuit implementing $\mathrm{exp}(i\tau\hat{H})$] for these mappings have not been discovered yet. In the relativistic case, the most convenient compact mapping is based on a subspace of symmetry-adapted functions employing the double group symmetry.

Assuming the NPA and the empty Dirac picture, the relativistic Hamiltonian has the same second quantized structure as the NR one

\vskip -0.3cm
\begin{equation}
  \label{ham}
  \hat{H} = \sum_{pq} h_{pq} a^{\dagger}_{p} a_{q} + \frac{1}{2}\sum_{pqrs} g_{pqrs} a^{\dagger}_{p} a^{\dagger}_{q} a_{s} a_{r}.
\end{equation}

\noindent
Here $h_{pq}$ and $g_{pqrs}$ denote one- and two-electron integrals that are in contrast to NR ones  in general complex. This is in fact no difficulty for a quantum computer, since our working environment is a complex vector space of qubits anyway and we do the exponential of a complex matrix even if the Hamiltonian is real (see Figure \ref{ipea_iteration}). After the decomposition of the unitary propagator [$\mathrm{exp}(i\tau\hat{H})$] to elementary quantum gates (in case of DM) using the Jordan-Wigner transform \cite{jordan_1928}, one can see that complex molecular integrals require twice as many gates compared to real ones \cite{whitfield_2010}, while complex arithmetic on a classical computer requires four times more operations.

The last of the aforementioned difficulties of the 4c formalism is the size of a Hamiltonian matrix eigenvalue problem. 
This can be inferred from the
observation that a significant larger number of integrals in the Hamiltonian
(\ref{ham}) will be non-zero due to the lowering of symmetry induced by spin-orbit
interaction. The loss of spin symmetry can to some extent be alleviated by
consideration of time reversal symmetry. In the Kramers-restricted (KR) approach
employed in this work the second-quantized Hamiltonian (\ref{ham}) is expressed  in
terms of a basis of Kramers pairs, that is, orbital pairs $\phi$ and
$\overline{\phi}$ connected by time reversal. Determinants may be
characterized by a pseudo-quantum number $M_K=1/2(N_A-N_B)$, reflecting the
different number of unbarred $N_A$ and barred $N_B$ bispinors. In the
non-relativistic limit the Kramers pairs can be aligned with spin partners
such that $M_K$ becomes identical to $M_S$. However, contrary to the NR limit,
determinants with different $M_K$ can mix in the presence of spin-orbit
interaction.
It can be shown (see Supplementary Information) that the ratio between dimensions of relativistic and non-relativistic Hamiltonian matrices scales as $\mathcal{O}(m^{1/2})$ in the number of molecular orbitals/bispinors.

When employing the DM on a quantum computer, this problem does not occur, since the Hamiltonian (\ref{ham}) then implicitly works with all possible values of $M_{K}$. The scaling of the relativistic qFCI method is therefore the same as the NR one, namely $\mathcal{O}(m^{5})$ \cite{lanyon_2010, whitfield_2010} , where $m$ is the number of molecular orbitals (bispinors). 

\begin{table}
  \begin{tabular}{c c c c}
    \hline
    \hline
    GAS & Min. el. & Max. el. & Shell types \\
    \hline
    I & 0 & 4 & $\sigma_{1/2}$, $\pi_{1/2}$ \\
    II & 2 & 4 & $\pi_{3/2}$ \\
    III & 4 & 4 & $\sigma_{1/2}^{*}$, 43 virtual Kramers pairs \\
    \hline
    \hline
  \end{tabular}
  \caption{GAS and occupation constraints for SbH $X~0^{+}$ and $A~1$ states CI calculations. The minimum and maximum number of electrons are accumulated values - apply to this and all preceding GA spaces.}
  \label{gas}
\end{table}

For numerical tests of the algorithm, we have chosen the SbH molecule whose non-relativistic ground state $^{3}\Sigma^{-}$ splits due to spin-orbit effects into $X~0^{+}$ and $A~1$. In the approximate $\lambda_\omega$ notation, these states are dominated by $\sigma_{1/2}^2\pi_{1/2}^2\pi_{3/2}^0$ and $\sigma_{1/2}^2\pi_{1/2}^1\pi_{3/2}^1$ configurations. The splitting is truly of ``molecular nature" as it disappears for dissociated atoms. Its experimental value is $\Delta E_{\rm{SO}} = 654.97$ cm$^{-1}$ \cite{balasubramanian_1989}. 

In all our simulations, we used the Dyall triple-zeta + valence correlating functions, total 28s 21p 15d 1f for Sb and cc-pVTZ (from EMSL basis set library) for H. \mbox{We, of course,} could not manage to simulate the FCI calculations with all electrons in such a large basis. We instead simulated general active space (GAS) KRCI computations \cite{fleig_2003} with the occupation constraints shown in Table \ref{gas} giving rise to CI spaces of approximately 29500 determinants. 
For a balanced description of both states, we optimized the spinors taking an average energy expression 
(2 electrons in 2 Kramers pairs $\pi_{1/2}$, $\pi_{3/2}$). 
We worked solely with a compact mapping employing the double-group symmetry ($C_{2v}^{*}$) and exponential of a Hamiltonian was simulated as an $n$-qubit gate (similarly as in \cite{aspuru-guzik_2005,wang_2008,veis_2010}). We used the DIRAC program \cite{dirac} for calculations of Hamiltonian matrices.
The nuclear potential $\phi_{nuc}$ was generated by finite nuclei using Gaussian charge distributions with exponents chosen according to Ref. \onlinecite{luuk:gnuc}. Simulations of qFCI computations were performed with our own C++ code \cite{veis_2010}. We ran 17 iterations of the IPEA with the difference between max. and min. expected energies equal to 0.5 $E_h$ We also did not count the least significant binary digit of the phase $\phi$ to the total success probability (for more details of the algorithm, we refer the reader to our preceding paper \cite{veis_2010}). This procedure corresponds to the final energy precision $\approx$$3.81 \times 10^{-6}$ $E_h$.

\begin{figure}
  \centering
  \includegraphics[width=7.5cm]{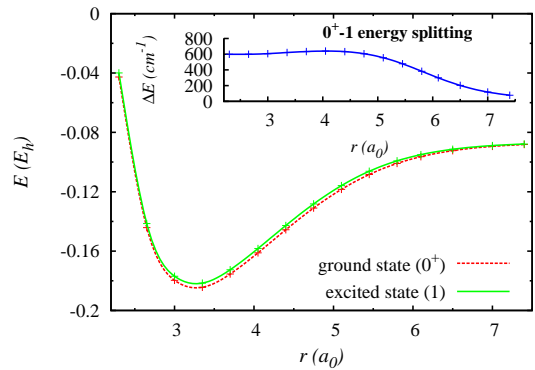}
  \caption{(Color online) Simulated potential energy curves of ground (0$^{+}$) and excited (1) states of SbH, and spin-orbit energy splitting. Absolute energies are shifted by $6481~E_h$.}
  \label{graf_en}
\end{figure}

\begin{figure}
  \centering
  \includegraphics[width=7.5cm]{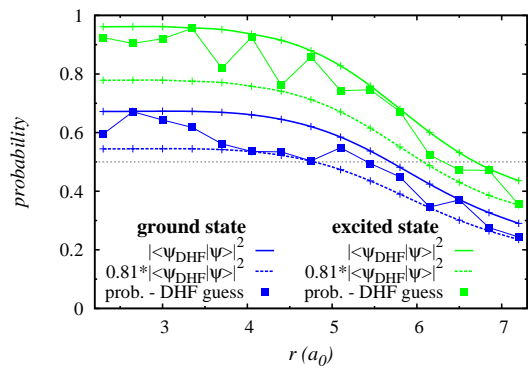}
  \caption{(Color online) SbH ground (0$^{+}$) and excited (1) state qFCI success probabilities (SPs) corresponding to HF initial guesses.}
  \label{prob}
\end{figure}

Simulated potential energy curves of both states are shown in Figure \ref{graf_en}. Based on our KRCI setup we obtain a vertical $\Delta E_{\rm{SO}}$ of 617 cm$^{-1}$. Success probabilities (SPs) of the algorithm with HF initial guesses ($\sigma_{1/2}^2 \pi_{1/2}^2 \pi_{3/2}^0$ for the $X~0^+$ state and $\sigma_{1/2}^2 \pi_{1/2}^1 \pi_{3/2}^1$ for $A~1$ one)
 are presented in Figure \ref{prob}. They correspond to the IPEA with the second part of a quantum register (encoding the relativistic quantum chemical wave function) maintained during all iterations (in \cite{veis_2010} denoted as version \textbf{A}). In this case, SPs always lie in the interval $|\langle \psi_{\rm{init}} | \psi_{\rm{exact}} \rangle |^{2} \cdot (0.81,1\rangle$ \cite{veis_2010}. Ground state SPs confirm that relativistic states have, due to near degeneracies caused by the spin-orbit coupling, often a stronger multireference character than non-relativistic ones. The upper bound of the SP is less than 0.7 even for the equilibrium geometry and HF initial guesses can in fact be safely used (SP $>$ 0.5, amplification of SP by repetitions) only up to 4.8 $a_0$. The SPs of the $A~1$ state are higher and HF initial guesses can be in a noise-free environment used up to 6 $a_0$.

The difficulty connected with a low success probability for the $X~0^{+}$ state at longer distances can be overcome by the ASP method \cite{aspuru-guzik_2005}. In this approach, one slowly varies the Hamiltonian of a quantum register, starting with a trivial one with a known eigenstate and ending with the final exact one in a following simple way 

\vskip -0.3cm
\begin{equation}
  \hat{H} = (1-s)\hat{H}_{\rm{init}} + s\hat{H}_{\rm{exact}} \quad s: 0 \rightarrow 1.
\end{equation}

\noindent
If the change is slow enough (depending on the gap between the ground and the first excited state), the register remains in its ground state according to the adiabatic theorem \cite{fahri_science_2001}. In our relativistic example, analogously to the non-relativistic one \cite{aspuru-guzik_2005}, $\hat{H}_{\rm{init}}$ is defined to have all matrix elements equal to zero, except $H_{11}$, which is equal to the (Dirac-)HF energy. 

\begin{figure}[!ht]
  \centering
  \includegraphics[width=7.5cm]{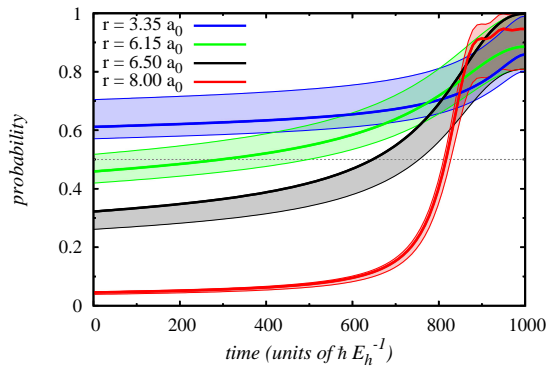}
  \caption{(Color online) Adiabatic state preparation (ASP) of the SbH ground state ($0^{+}$) for different internuclear distances. Solid lines correspond to qFCI success probabilities, $|\langle \psi_{\rm{ASP}} | \psi_{\rm{exact}} \rangle |^{2} \cdot (0.81,1\rangle$ interval is colored. 1000 $\hbar E_{h}^{-1}$ $\approx 10^{-14}$ s.}
  \label{asp}
\end{figure}

\begin{figure*}[!ht]
  \begin{center}
  \mbox{
    \Qcircuit @C=0.4em @R=0.4em {
    & \qw & \qw & \qw & \qw & \qw & \qw & \qw & \qw & \qw & \qw & \qw & \gate{R_z} & \targ & \gate{R_z}  & \targ & \gate{R_z} & \targ & \gate{R_z} & \targ & \qw & \qw & \qw & \qw & \qw & \qw & \qw & \qw \\
    & \gate{S} & \qw & \targ & \gate{\color{white}{\rule{0.2cm}{0.2cm}}} & \targ & \qw & \gate{S^{\dagger}} & \ctrl{1} & \qw & \ctrl{1} &\qw & \gate{R_{z}} & \ctrl{-1} \qw & \qw & \qw & \qw & \ctrl{-1} & \qw & \qw & \gate{S} & \qw & \targ & \gate{\color{white}{\rule{0.2cm}{0.2cm}}} & \targ & \qw & \gate{S^{\dagger}} & \qw \\
    & \gate{S} & \gate{H} & \ctrl{-1} & \gate{\color{white}{\rule{0.2cm}{0.2cm}}} & \ctrl{-1} & \gate{H} & \gate{S^{\dagger}}  & \targ & \gate{R_{z}} & \targ & \gate{R_{z}} & \gate{R_{z}} & \qw & \qw & \ctrl{-2} & \qw & \qw & \qw & \ctrl{-2} & \gate{S} & \gate{H} & \ctrl{-1} & \gate{\color{white}{\rule{0.2cm}{0.2cm}}} & \ctrl{-1} & \gate{H} & \gate{S^{\dagger}} & \qw  
  }}
  \end{center}
  \caption{Scheme of a circuit corresponding to CAS(4,3) calculations on SbH. Empty squares represent generic single-qubit gates. $R_{z}$ gates are without angle specification. For derivation, details, and all the parameters, see Supplementary Information.}
  \label{circuit_scheme}
  \vspace{-0.4cm}
\end{figure*}
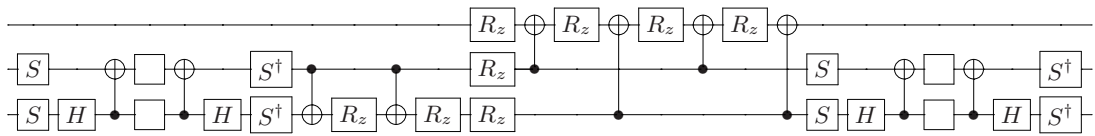

We simulated $X~0^{+}$ qFCI computations with adiabatically prepared states for different internuclear distances; results are shown in Figure \ref{asp}. In this case, for computational reasons, we employed complete active space (CAS) KRCI method with a CAS composed of 2 electrons in the highest occupied ($\pi_{1/2}$) and 45 lowest unoccupied Kramers pairs (corresponds to 2116 determinants). It can be seen that for $t = 1000$ $\hbar E_{h}^{-1}$, the upper bound of the SP goes safely to unity even for $r = 8$ $a_0$. 

Recently, there appeared two papers presenting the first physical implementations of \textit{non-relativistic} 
qFCI computations on optical \cite{lanyon_2010} and NMR \cite{du_2010} quantum computers. Correspondingly, we would like to propose two candidates for the first \textit{relativistic} computations on real quantum computers. Our proposals represent ``digital (circuit-based) quantum simulations" (DQS) as defined by Bulata and Nori \cite{nori_science_2009}. Conceptually different are ``analogue quantum simulations" (AQS), where the evolution of a studied quantum system is mapped to be simulated onto the controlled evolution of the quantum simulator. Recently, Gerritsma et al. used this approach for the proof-of-principle simulation of a one-dimensional Dirac equation with a single trapped ion \cite{gerritsma_2010}.

Both of our examples represent calculations of SbH $^{3}\Sigma^{-}$ ground state spin-orbit splitting. Since one has to employ rather large basis sets (triple-$\zeta$ quality) to get a meaningful result, they again are not true FCI calculations, but FCI calculations in a limited CAS. The first one corresponds to a CAS composed of 2 electrons in the highest occupied ($\pi_{1/2}$) and the lowest unoccupied ($\pi_{3/2}$) Kramers pairs [CAS(2,2)]. After the factorization of a Hamiltonian according to the $\Omega$ quantum number and taking into account only one of the two degenerate $z$-projections of $\Omega$ (for $\Omega = 1$), the size of the CI space is 2 for the ground state (0$^{+}$) and 1 for the excited state (1). The excited state is therefore trivial and the calculation of the ground state is in fact a complete analogue of the already mentioned NR computations \cite{lanyon_2010,du_2010}, because it needs just one qubit for the wave function (2 in total). The controlled single-qubit gate can be decomposed using 2 controlled NOTs (CNOTs) \cite{nielsen_chuang}. Calculations with this active space yield an $\Delta E_{\rm{SO}} = 509$ cm$^{-1}$ computed at the experimental equilibrium bond distance of 3.255 $a_0$.

The second example represents a 3-qubit experiment (2 qubits for the wave function) and employs a CAS composed of 4 electrons in the $\sigma_{1/2}\pi_{1/2}\pi_{3/2}$ Kramers pairs [CAS(4,3)]. It gives a better value of $\Delta E_{\rm{SO}} (518$ cm$^{-1}$) than CAS(2,3). After $\Omega$ factorization, the CI space of the excited state has a dimension 3 and that of the ground state 5. Fortunately, near the equilibrium bond distance, the Hamiltonian matrix of the ground state is to a very good approximation block diagonal (ground state energy difference of the order $\mu E_{\rm{h}}$), coupling only 3 configurations ($\sigma_{1/2}^{2} \pi_{1/2}^{2} \pi_{3/2}^{0}$, $\sigma_{1/2}^{2} \pi_{1/2}^{0} \pi_{3/2}^{2}$, and $\sigma_{1/2}^{0} \pi_{1/2}^{2} \pi_{3/2}^{2}$). If we take into account only these configurations, both states can be encoded by two qubits. 

We used the \textit{Quantum Shannon Decomposition} (QSD) technique \cite{shende_2006} and decomposed the controlled action of a two-qubit $\mathrm{exp}(i\tau\hat{H})$. QSD is known to decompose a generic three-qubit gate with the least number of CNOTs (20). A minimal number of CNOTs is very important as their implementations are orders of magnitude more difficult. We found a circuit with 9 CNOTs which is not universal in the sense that the decomposition must be done for all powers of U individually, or a universal 10-CNOT-circuit.  The structure of this circuit is shown in Figure \ref{circuit_scheme}.
The controlled action of $n$th power of $U$ is simply done by multiplication of the angles of $R_{z}$ rotations by $n$. Details of the decomposition and also all parameters important for a possible experimental realization which correspond to the calculations at internuclear distance 3.255 $a_0$ can be found in the Supplementary Information. The proposed experiments are undoubtedly a challenge for different realizations of quantum computation. We regard experimental verification of the usage of HF initial guesses in a realistic noisy environment and also the performance of both versions of IPEA (\textbf{A} and \textbf{B})  proposed in \cite{veis_2010} as very interesting.

\textit{Conclusion. -} In this work, we have presented the first quantum algorithm for 4c relativistic FCI energy computations. This algorithm not only achieves an exponential speedup over its classical counterpart, but also has the same cost (in terms of scaling) as its NR analogue. We have proved its functionality by numerical simulations of calculations of the spin-orbit splitting in SbH. We have also proposed and designed the first small-scale experimental realizations of relativistic qFCI computations. Our algorithm can be used stand-alone or as a subroutine of a property algorithm of Kassal et. al. \cite{kassal_2009} e.g. for calculations of NMR properties. 

This work has been supported by the GA\v{C}R (203/08/0626) and the GAUK (114310).
Lu.V. has been supported by NWO through the VICI programme. S.K. acknowledges a postdoctoral grant from FNU.

\clearpage

\renewcommand{\theequation}{S\arabic{equation}}
\renewcommand{\thetable}{S\Roman{table}}
\setcounter{equation}{0} 
\setcounter{table}{0} 
\section{Supplementary Information}

\subsection{Relativistic Hamiltonian approximations}
Our work is based on the 4-component electronic Dirac-Coulomb Hamiltonian which in atomic units is given as

\begin{equation}
  \hat{H} = \sum_{i=1}^N \left [c(\boldsymbol\alpha_i \cdot \mathbf{p}_i) + \beta^{\prime}_i m c^2 - \phi_{nuc} \right]
+ \sum_{i<j} \frac{1}{r_{ij}} +V_{NN}.
\end{equation} 

\noindent
We work within the Born-Oppenheimer clamped nuclei approximation which allows to factorize out time-dependence of the one-electron
problem in the nuclear frame. The one-electron operator of the electronic Hamiltonian is accordingly given by the Dirac Hamiltonian
in the electrostatic potential $\phi_{nuc}$ of clamped nuclei. The relativistic energy scale has been aligned with the non-relativistic one by subtraction of the electron rest mass.

The full Lorentz-invariant two-electron interaction can not be written 
in a simple closed form, so approximation and thus loss of strict Lorentz invariance is in practice unavoidable.
In Coulomb gauge the zeroth-order $\mathcal{O}(c^{0})$ operator is given by the Coulomb term employed here.
This resulting Hamiltonian covers the major part of the spin-orbit interaction, including two-electron spin-same orbit, as well as scalar relativistic effects. Experience 
shows that the Coulomb term is enough for most chemical purposes \cite{visser:hyd4}, but for highly accurate molecular spectra the Breit (Gaunt) term, carrying 
spin-other orbit interaction, is recommended.

A fundamental conceptual problem is that the Dirac-Coulomb(-Breit) Hamiltonian has no bound solutions
due to the one-electron negative-energy continuum solutions generated by the Dirac Hamiltonian \cite{brown:ravenhall}. We adopt the no-pair approximation (NPA),
widely used in relativistic quantum chemistry \cite{dyall}, in which the N-particle basis of Slater determinants is constructed from positive-energy bispinors only. This procedure in fact neglects all QED effects, but it is justifiable at the energy scale relevant to chemistry. In particular, the Born-Oppenheimer approximation is expected to have larger impact than the neglect of QED effects. 

We finally note that the Fock space approach to include positronic states within the Dirac-Coulomb(-Breit) Hamiltonian approximation \cite{saue:wilson,Kutzelnigg_CP2011} should be tractable on a quantum computer as well, since the direct mapping (including qubits for positrons) covers the whole Fock space generated by a finite basis set.
For further discussion of the Dirac-Coulomb approximation and how to possibly go beyond it the reader may consult Refs.\cite{saue:wilson,saue:hamprimer,Kutzelnigg_CP2011,liu:PCCP2012,derezinski2012}.

\subsection{Size of 4c relativistic FCI eigenvalue problem}
In this section, we compare dimensions of non-relativistic and 4c relativistic Hamiltonian matrices. In the NR case, the Hamiltonian matrix is block diagonal according to $M_{S}$. Thus for a closed shell system with $n$ electrons in $m$ orbitals, the number of determinants is

\begin{equation}
  N_{\rm{NR}} = \left( \begin{array}{c}  m \\ n/2 \end{array} \right)^2 .
\end{equation}

\noindent
The relativistic Hamiltonian mixes determinants with different $M_{K}$ values and therefore

\begin{equation}
  N_{\rm{R}} = \left( \begin{array}{c}  2m \\ n \end{array} \right).
\end{equation}

\noindent
Using Stirling's approximation in the form

\begin{equation}
  \mathrm{ln}~m! \approx \frac{1}{2} \mathrm{ln}~(2\pi m) + m\mathrm{ln}~m - m \qquad \mathrm{for}~m\rightarrow \infty,
\end{equation}

\noindent
and setting $m = k \cdot n$, the ratio between the relativistic and non-relativistic number of determinants is given by the expression

\begin{equation}
  k_{\rm{R}/\rm{NR}} = \frac{N_{\rm{R}}}{N_{\rm{NR}}} = \Bigg( \frac{\sqrt{\pi (2k - 1)}}{2k}\Bigg) \cdot m^{1/2}.
\end{equation}

\subsection{Controlled-U circuit design}

In this section, we construct a quantum circuit which corresponds to the controlled action of powers of $U=e^{i\tau\hat{H}}$ (see Figure 1 of the paper) for a CI space of dimension 3. For this case, we need two qubits to encode the quantum chemical wave function and $U$ has a block diagonal structure with $3 \times 3$ block of an exponential of a Hamiltonian and unity on a diagonal to complete the vector space of two qubits.

We use the \textit{Quantum Shannon Decomposition} technique of Shende et. al. \cite{shende_2006}. It turns out to be very useful to generalize the concept of controlled gates to quantum multiplexors. A quantum multiplexor is a quantum conditional which acts on target qubit(s) in a different way, according to the state of select qubit(s). If the select qubit is the most significant one, then it has the following matrix form

\hskip 1cm
\begin{minipage}{0.08\textwidth}
\begin{center}  
\vskip 0.3cm 
  \mbox{
  \Qcircuit @C=0.8em @R=0.8em {
    & \ctrlm{1} & \qw \\
    & \gate{U} & \qw
  }
}
\end{center}
\end{minipage}
\hskip -1.2cm
\begin{minipage}{0.39\textwidth}
  \begin{equation}
    = \qquad \left( \begin{array}{cc}  U_0 & ~ \\ ~ & U_1 \end{array} \right).
  \end{equation}
\end{minipage}
\vskip 0.4cm

\noindent
It performs $U_0$ on the target qubit if the select qubit is $\ket{0}$ and $U_1$ if the select qubit is $\ket{1}$. A controlled gate is a special case where $U_0 = I$. More generally, if $U$ is a quantum multiplexor with $s$ select qubits and $t$ target qubits and the select qubits are most significant, the matrix of $U$ will be block diagonal, with $2^s$ blocks of size $2^{t} \times 2^{t}$.

A controlled 2-qubit $U$ (c-$U_{2q}$) is a special case of multiplexed $U$ and can be decomposed in the following way \cite{shende_2006}

\begin{minipage}{0.10\textwidth}
\begin{center}  
\vskip 0.70cm 
  \mbox{
  \Qcircuit @C=0.8em @R=0.8em {
    & \ctrl{1} & \qw \\
    & \multigate{1}{U} & \qw \\
    & \ghost{U} & \qw
  } 
  }
\end{center}
\end{minipage}
\begin{minipage}{0.08\textwidth}
  \vskip 0.6cm
  \begin{center} = \end{center}
\end{minipage}
\hskip -1.3cm
\begin{minipage}{0.35\textwidth}
  \begin{equation}
  \Qcircuit @C=0.8em @R=0.5em {
    & \qw & \gate{R_z} & \qw & \qw \\
    & \multigate{1}{W} & \ctrlm{-1} & \multigate{1}{V} & \qw \\
    & \ghost{W} & \ctrlm{-1} & \ghost{V} & \qw
  }
\label{cu}
\end{equation}
\end{minipage}
\vskip 0.4cm

\noindent
A multiplexed $z$-rotation in the middle of the circuit on the right-hand side (at this stage without angle specification) is in fact a diagonal matrix with second half of a diagonal equal to a Hermitian conjugate of the first one.
The circuit equation (\ref{cu}) corresponds to the matrix equation 

\begin{equation}
  \left( \begin{array}{cc}  I & ~ \\ ~ & U \end{array} \right) = \left( \begin{array}{cc}  V & ~ \\ ~ & V \end{array} \right) \left( \begin{array}{cc}  D & ~ \\ ~ & D^{\dagger} \end{array} \right) \left( \begin{array}{cc}  W & ~ \\ ~ & W \end{array} \right).
\end{equation}

\noindent
Note that right in the equation means left in the circuit as the time in a circuit flows from the left to the right. 

We then have

\begin{eqnarray}
  \label{w}
  I & = & V D W, \\
  U & = & V D^{\dagger} W, \\
  \label{u_diag}
  U^{\dagger} & = & V D^{2} V^{\dagger}.
\end{eqnarray}

A single-multiplexed $R_z$ gate (with angle $\phi_0$ for $\ket{0}$ state of a select qubit and $\phi_1$ for $\ket{1}$) can be implemented with the following circuit

\vskip 0.3cm
\hskip -0.5cm
\begin{minipage}{0.08\textwidth}
  \mbox{
  \Qcircuit @C=0.8em @R=0.8em {
    & \ctrlm{1} & \qw \\
    & \gate{R_z} & \qw
  }
  }
\end{minipage}
\begin{minipage}{0.03\textwidth}
  =
\end{minipage}
\hskip -0.5cm 
\begin{minipage}{0.40\textwidth}
  \vskip -0.5cm
  \begin{equation}
  \Qcircuit @C=0.8em @R=0.5em {
    & \qw & \ctrl{1} & \qw & \ctrl{1} & \qw \\
    & \gate{R_{z}(\frac{\phi_{0} + \phi_{1}}{2})} & \targ & \gate{R_{z}(\frac{\phi_{0} - \phi_{1}}{2})} & \targ & \qw 
  }
  \hskip 0.2cm ,
  \end{equation}
\end{minipage}
\vskip 0.4cm

\noindent
since $\sigma_x$ gates on both sides of $R_z$ turn over the direction of the $R_z$ rotation. If we use this approach for demultiplexing the $R_{z}$ gate in (\ref{cu}), we end up (after some simple circuit manipulations) with the following circuit for c-$U_{2q}$

\begin{equation}
\label{circuit}
\begin{small}
\Qcircuit @C=0.3em @R=0.4em {
  & \gate{R_z(\varphi_1)} & \targ & \gate{R_z(\varphi_2)}  & \targ & \gate{R_z(\varphi_3)} & \targ & \gate{R_z(\varphi_4)} & \targ & \qw & \qw \\
  & \multigate{1}{W} & \ctrl{-1} \qw & \qw & \qw & \qw & \ctrl{-1} & \qw & \qw & \multigate{1}{V} & \qw \\
  & \ghost{W}  & \qw & \qw & \ctrl{-2} & \qw & \qw & \qw & \ctrl{-2} & \ghost{V} & \qw  
}
\end{small}
\end{equation}

\noindent
where

\begin{eqnarray}
  \label{phi}
  \varphi_1 & = & \frac{1}{4}(\phi_{00} + \phi_{01} + \phi_{10} + \phi_{11}), \\
  \varphi_2 & = & \frac{1}{4}(\phi_{00} + \phi_{01} - \phi_{10} - \phi_{11}), \nonumber \\
  \varphi_3 & = & \frac{1}{4}(\phi_{00} - \phi_{01} - \phi_{10} + \phi_{11}), \nonumber \\
  \varphi_4 & = & \frac{1}{4}(\phi_{00} - \phi_{01} + \phi_{10} - \phi_{11}). \nonumber 
\end{eqnarray}

\noindent
Individual $\phi$'s in (\ref{phi}) can be extracted from the diagonal of $D$, which has the form: diag($e^{-i\phi_{00}}$,$e^{-i\phi_{01}}$,$e^{-i\phi_{10}}$,$e^{-i\phi_{11}}$). 

We would like to emphasize that this is not intended to be a decomposition technique for general $U$'s, as it itself requires classical diagonalization [of $U^{\dagger}$, see (\ref{u_diag})]. A general \textit{efficient} decomposition of an exponential of a Hamiltonian to elementary gates is known only for the direct mapping \cite{lanyon_2010, whitfield_2010}. But this mapping is not suitable for small scale experiments due to the relatively high number of required qubits and operations thereon. Our aim was in fact to prepare the ground for a first \textit{non-trivial} (more than one qubit in the quantum chemical part of the register) experimental realization of (relativistic) quantum chemical computation on a quantum computer.

Because $V$ belongs to the group \textbf{O}(4) (matrix of eigenvectors of a symmetric matrix), it can be decomposed using only two CNOT gates \cite{vatan_2004}:

\begin{equation}
  \label{v_circuit}
  \Qcircuit @C=0.8em @R=0.4em {
    & \gate{S} & \qw & \targ & \qswap & \gate{A} & \targ & \qw & \gate{S^{\dagger}} & \qw \\
    & \gate{S} & \gate{H} & \ctrl{-1} & \qswap \qwx & \gate{B} & \ctrl{-1} & \gate{H} & \gate{S^{\dagger}} & \qw \gategroup{1}{5}{2}{5}{.7em}{--}
  }
\end{equation}

\vskip 0.1cm
\noindent
$H$ and $S$ are standard Hadamard and phase gates and $A$, $B$ are generic single-qubit gates that can be further decomposed e.g. by $Z$-$Y$ decomposition \cite{nielsen_chuang}


\begin{equation}
  \label{zydec}
  A = e^{i\alpha} R_{z}(\beta) R_{y}(\gamma) R_{z}(\delta).
\end{equation}

\noindent
There is a highlighted swap gate in (\ref{v_circuit}) which should be applied only if the determinant of $V$ is equal to $-1$ \cite{vatan_2004}. 

The matrix $W$, on the other hand, is not real as it is equal to $D^{\dagger}V^{\dagger}$ (\ref{w}) and can be implemented using three CNOT gates (see e.g. \cite{vatan_2004,shende_2004}). The total count is thus 9 CNOTs.

The disadvantage of the aforementioned scheme is that $W$ must be decomposed for each power of $U$ individually. If we separate $W$ to $V^{\dagger}$ and $D^{\dagger}$, $V^{\dagger}$ is the same for all powers of $U$ (eigenvectors don't change) and $D^{\dagger}$ can be up to a non-measurable global phase implemented with the following circuit  

\begin{equation}
  \label{d_circuit}
  \Qcircuit @C=0.8em @R=0.4em {
    & \ctrl{1} & \qw & \ctrl{1} &\qw & \gate{R_{z}(\varphi_{6})} & \qw \\
    & \targ & \gate{R_{z}(-\frac{\varphi_{5}}{2})} & \targ & \gate{R_{z}(\frac{\varphi_{5}}{2})} & \gate{R_{z}(\varphi_{7})} & \qw
  }
\end{equation}

\noindent
where

\begin{eqnarray}
  \label{phi2}
  \varphi_{5} & = & \frac{1}{2}(\phi_{00} - \phi_{01} - \phi_{10} + \phi_{11}), \nonumber \\
  \varphi_{6} & = & \frac{1}{4}(-\phi_{00} - \phi_{01} + \phi_{10} + \phi_{11}), \\
  \varphi_{7} & = & \frac{1}{2}(-\phi_{00} + \phi_{01}). \nonumber 
\end{eqnarray}

\noindent
The circuit for $V^{\dagger}$ is the same as for $V$ (\ref{v_circuit}), merely $A$ is replaced by $B^{\dagger}$ and $B$ by $A^{\dagger}$.

Presented 10-CNOT-circuit is universal for all powers of $U$. The only thing one has to do is to multiply the angles of $R_{z}$ rotations in (\ref{circuit}) and (\ref{d_circuit}) according to the power of $U$, e.g. by 2 for the second power.

Table \ref{par} summarizes the circuit parameters for ground as well as excited state calculations described in the preceding text. Notice that $\phi_{11}$ is zero in both cases by construction. To complete the vector space of two qubits, we in fact added one eigenvalue of the Hamiltonian equal to zero. Other simplification, which originates from the block diagonal structure of $U$, is that $A$ and $B$ matrices in the decomposition of $V$ (\ref{v_circuit}) differ only in a global phase. Because the global phase is not measurable, we present just the angles of rotations. Also only the parameters corresponding to $A$ and $B$ are shown. Going to their Hermitian conjugates means swapping of $\beta$ and $\delta$ and changing the sign of all of them.

\begin{table}[t]
  \begin{tabular}{c c c}
    \hline 
    \hline
     & Ground state ($0^{+}$) & Excited state (1) \\
    \hline
    $\phi_{00}$ & -1.01642278 & -1.00656763 \\
    $\phi_{01}$ & -0.68574813 & -0.18597924 \\
    $\phi_{10}$ & 0.69657237 & -0.39129153 \\
    $\phi_{11}$ & 0 & 0 \\
    \hline
    $\beta$ & 0.73125768 & -0.00680941 \\
    $\gamma$ & -0.10311594 & 2.21832498 \\
    $\delta$ & -0.12107336 & -3.13494247 \\
    \hline
    $\Delta E_{\rm{shift}}$ & -6477.89247780 & -6477.89247780 \\
    \hline 
    \hline
  \end{tabular}
  \caption{Circuit parameters: rotation angles $\phi_{ij}$, $i,j \in \{0,1\}$ (\ref{phi},\ref{phi2}), $Z$-$Y$ decomposition parameters of $A$, $B$ (\ref{v_circuit}) and energy shifts (core energy + nuclear repulsion) for CAS(4,3) calculations of $0^{+}$ and $1$ states. For the details see preceding text.}
  \label{par}
\end{table}

For the excited state, the determinant of $V$ is equal to $-1$ and therefore the swap gate in (\ref{v_circuit}) should be applied. Because we took Hamiltonian matrices from the DIRAC program \cite{dirac}, the parameters in Table \ref{par} refer to the difference between the total energy and core energy + nuclear repulsion ($\Delta E_{\rm{shift}}$). The presented method with the parameters form Table \ref{par} implements the exponential $e^{i\tau\hat{H}}$, as was already mentioned. But in our version of the algorithm \cite{veis_2010}, we in fact need $e^{-i\tau\hat{H}}$. The obtained energy therefore corresponds to the negative of the energy. For the negative, the energy guesses $E_{\mathrm{max}} = 3.5$ and $E_{\mathrm{min}} = 2.0$ corresponding to the maximum and minimum expected energies were used.

We don't give any explicit proof that the \textit{Quantum Shannon decomposition} is optimal in the number of CNOT gates for the specific case of block diagonal c-$U_{2q}$. However, this conjecture is supported by the fact that we also implemented the Group Leaders Optimization Algorithm (GLOA) of Dashkin and Kais \cite{daskin_2011} and unsuccessfully tried to find a better circuit (in terms of number of controlled operations) with a fidelity error smaller than 0.01.

\bibliographystyle{h-physrev}
\bibliography{kvantove_pocitace}

\end{document}